\documentclass[preprint,eqsecnum,aps]{revtex4}
\usepackage{dcolumn}
\usepackage{graphicx}
\begin{document}
\title{A novel braneworld model with a bulk scalar field}
\author{
Ratna Koley \footnote{Electronic address : {\em ratna@cts.iitkgp.ernet.in}}${}^{}$ 
and Sayan Kar \footnote{Electronic address : {\em sayan@cts.iitkgp.ernet.in}
}${}^{}$}
\affiliation{Department of Physics and Meteorology and Centre for
Theoretical Studies \\Indian Institute of Technology, Kharagpur 721 302, India}
%\date{\today}
%\maketitle
%\twocolumn[
%\widetext
%\abstract

\begin{abstract}
We consider a new braneworld model with a bulk scalar field
coupled to gravity. The bulk scalar action is inspired
by the proposed low energy effective action around the tachyon vacuum.  
A class of warped geometries representing
solutions of this Einstein-scalar system for a specific scalar potential
is found. The geometry is non singular with a decaying
warp factor and a negative Ricci curvature. The solution of the
hierarchy problem is obtained using this type of warping. Though
qualitatively similar to the usual Randall--Sundrum I model
there are interesting quantitative differences.
Additionally, in the RS-II set up the graviton zero mode as well as spin half
massless fermions are found to be localised on the brane.      
\end{abstract}
%]
\maketitle

\section{Introduction}  

The study of extra dimensions (D $>$ 4) has been of considerable interest 
to theoretical physicists ever since Kaluza-Klein introduced them
in their attempt to unify gravity with electromagnetism {\cite{kk}}. 
Over the last few decades, the notion of extra dimensions
has been discussed with renewed enthusiasm in the 
context of string theory {\cite{string}} where their presence is
inevitable. 
More recently, attention has shifted toward the
{\em braneworld} picture - in which our four dimensional universe
(3-brane) is viewed as an embedded hypersurface in a higher
dimensional bulk spacetime. In the now-popular Randall-Sundrum models 
(known as RS I and RS II) {\cite{rs}}
of warped (nonfactorisable) spacetimes the extra dimensions exist, can be
compact or non-compact, but the usual schemes of {\em compactification}
are replaced by the notion of {\em localisation} of fields on the brane. 
The RS-I model provides us with a solution of the gauge
hierarchy problem whereas RS-II shows us how the higher dimensional gravity
can still reproduce Newtonian gravity on the brane. Several problems
exist with the RS models, some of which have been tackled by 
introducing appropriate modifications {\cite{modrs}. 

In the RS models, the bulk is essentially vacuum but with a negative 
(five dimensional) cosmological constant (AdS spacetime). 
This leads to a problem of stability of the branes -- a matter of concern 
in the two--brane (RS I) scenario. To remedy this, bulk fields are introduced,
in particular, a scalar field with a potential {\cite{radion}}. 
A bulk scalar also provides us with a way of generating the braneworld as a domain wall in
five dimensions--these are the so--called {\em thick} branes. 
The last few years has seen much
work being carried out on diverse aspects of such braneworld models
with variety of bulk field configurations {\cite{rs,thick}}. 
In the context of such newer models with bulk fields, issues such as
localisation of various types of matter fields {\cite{loc,gravloc,fermiloc}}, 
cosmological consequences {\cite{cosmo}} and
particle phenomenology in the braneworld context {\cite{particle}} have been
addressed to some extent. Further
consequences of new types of bulk matter and their resulting effects still 
remains an open arena of research.     

Recently there has been a lot of activity on a newly proposed
scalar field theory - namely `tachyon matter' arising in the context
of string theory {\cite{sen}}. Briefly, one can say that open string
tachyons (attached to D--branes and reflecting D--brane instability), 
residing on the unstable (tachyonic) maximum of the tachyon potential can
acquire a vacuum expectation value (condensate) and roll down to a stable
configuration. The energy of such a tachyon condensate has been
calculated in string field theory.     
The effective action thus obtained is
somewhat uncommon due to its special form. It is completely different from
the usual scalar field action and resembles, in some sense, 
a Born--Infeld type action {\cite{bi}}. 
Though proposed by several authors independently 
{\cite{tachyon}} its physical consequences have been
analysed in greater detail recently by Sen {\cite{sen}}. It has been shown 
that such a scalar field might play a 
role in cosmology in the context of dark matter/energy. 
This is because the effective energy--momentum tensor can be shown to be
{\em equivalent} to that 
of noninteracting, nonrotating dust {\cite{tachycosmo}}.   
Since, currently, there is no unique, experimentally verifiable choice 
for bulk matter, it is surely useful to explore various possibilities
{\cite{newbulk}}. 
This has motivated us to consider the tachyon condensate as a
bulk field coupled to gravity and study its effects on 
the geometry. It is worth mentioning here that one might also visualise the
tachyon condensate action as a new type of scalar field action
without making any particular reference to string theory as such. The
dynamics of this scalar field, both classical and quantum is of interest
in its own right and has been investigated from this standpoint too.  

The bulk geometry is taken to be non-factorisable (warped). We first obtain 
an exact solution of the full set of Einstein-scalar equations with a given
scalar potential. Thereafter, we address the hierarchy problem in a RSI set-up. 
In the subsequent section, we analyse the graviton zero mode and demonstrate
its localisation in RSII set-up. Finally, we concentrate on fermion (massless)
localisation and conclude with a summary of the results obtained.  

\section {The model and the exact solution}

Let us begin with the action for the bulk scalar field {\cite{sen}} given by 

\begin{equation}
S_{T} = \alpha_{T} \int d^{5}x \sqrt{-g} V(T)\sqrt{1+g^{MN}\partial_{M} T\partial_{N} T}
\end{equation}

where $\alpha_{T}$ is an arbitrary constant, $g_{MN}$ being the
five dimensional metric. The scalar field is represented by T and 
V(T) corresponds to its potential. The constant $\alpha_{T}$ can take
either positive or negative values. 
The energy momentum tensor corresponding to the above action is given
by :

\begin{equation}
T_{MN}^{(T)} = \frac{\alpha_{T}}{2} \left[ g_{MN} V(T) \sqrt{1 +
    (\nabla T)^2} - \frac {V(T)}{\sqrt{1+ (\nabla T)^2}} 
\partial_{M} T \partial_{N} T \right] 
\end{equation}

For an appropriate choice of
$\alpha_{T}$ and V(T) it is possible to keep the geometry, produced by
the above action along with gravity, unaltered
with respect to the signature change of $\alpha_{T}$.
Note the unusual form of the action - the potential is
multiplied with a square root function containing the metric and derivatives
of the scalar field. 
This inherent novelty of the action has generated a lot of activity. 
The full five dimensional action will also contain an Einstein--Hilbert term
along with the contributions from the brane vacuum energies. 

The action of the model is
\begin{eqnarray}  
S & = & S_{G} + S_{T} + S_{B} \\
S_{G} & = &
%{\frac{1}{16 \pi G_{5}}} 
2 M^3 \int \sqrt{-g} R d^{5} x \\
S_{B} & = & - \int  \sum_{j}  \tau_{(j)} \sqrt{-\tilde{g}^{(j)}} d^{4} x 
\end{eqnarray}

where $\tilde{g}_{\mu\nu}^{(j)}$ is the induced metric on the brane and
$\tau_{(j)}$ is the vacuum energy density of the brane located at $\sigma =
\sigma_{j}$. There are two branes in RSI set-up and a single brane in
RSII set-up.  
Variation of the total action w.r.t. 
$g_{MN}$ leads to the Einstein--scalar equation and that w.r.t. T 
gives rise to the scalar field equation.

We assume a line element, in the five dimensional bulk, of the form: 
\begin{equation}
ds^2 = e^{-2 f(\sigma)}(\eta_{\mu\nu} dx^{\mu}dx^{\nu}) + d\sigma^2
\end{equation}

The warp factor f($\sigma$) and the scalar field T are 
considered to be function of $\sigma$ only. With these
ansatz, the Einstein-scalar equation reduces to the following system of
coupled, nonlinear ordinary differential equations :

\begin{eqnarray}
{f'}^2 & = & \frac{a}{2} \frac{V(T)}{\sqrt{1 + T'^2}} 
\\
f '' & = & - a V(T) \frac{T'^2}{\sqrt{1 + T'^2}} + \frac {1}{12 M^3} \sum_{j}\tau_{j}
\delta(\sigma - \sigma_{j}) 
\end{eqnarray}

where $a =  \frac{\alpha_{T}}{12 M^3}$ and prime denotes derivative with
respect to the coordinate $\sigma$. 
We have been able to find an exact analytical solution of 
the above set of equations (2.7) - (2.8) for the scalar field potential 

\begin{equation}
V(T) = \frac{8}{a k^2} \frac{1}{T^4} \left (1 +
    \frac{ k^2}{4} T^2 \right)^{\frac{1}{2}}
\end{equation} 

where $k$ is an arbitrary constant. It is straightforward to show that
the potential remains non-singular as long as the scalar field does not
become zero. We will show that in our set of exact solutions the
scalar field T, never becomes zero for any finite value of the extra
coordinate. It is worth noting that our solution is for $\alpha_T >0$.
If we were to take $\alpha_T <0$ (which is the usual choice, $\alpha_T = -1$
in the literature on tachyon matter) the same solution will be valid
provided we invert the potential by absorbing the minus sign in the
definition of $V(T)$. Note that with $a\rightarrow -a$ and $V(T)\rightarrow
-V(T)$ the field equations remain unchanged.

\begin{figure}[htb]
\includegraphics[width= 8cm,height=5cm]{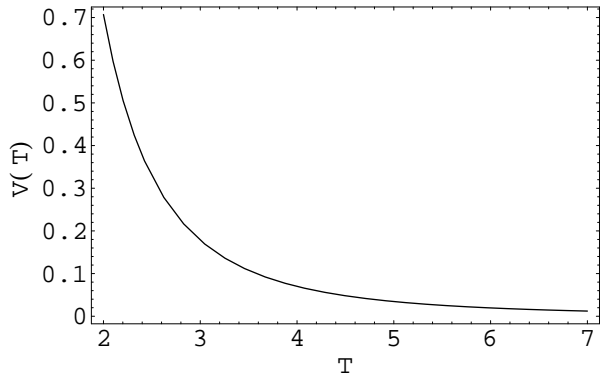}
\caption{The scalar field potential is depicted in the above figure
 for the parameters $a_{1} = k = 1$ and $\alpha_{T}$ having unit
 positive value} 
\end{figure}

The explicit forms of the warp factor and the scalar field are given by :
\begin{eqnarray}
f(\sigma) = - \frac{a_{1}}{k} e^{- k \vert \sigma \vert} \\
T(\sigma) = \sqrt{\frac{2}{k a_{1}}} e^{\frac{k}{2} \vert \sigma \vert}
\end{eqnarray}

for the entire domain of the extra dimension, $- \infty <\sigma < \infty $. 
Where $a_{1}$ is an arbitrary constant and $k$
is a scale of the order of Planck scale. It is important to note that
the metric warp factor is a super exponential function which decays
as one moves along the transverse dimension. The scalar is an exponentially
growing function of $\sigma$. Also, the scalar
field never becomes zero, keeping the potential always
non-singular.

\begin{figure}[htb]
\includegraphics[width= 8cm,height=5cm]{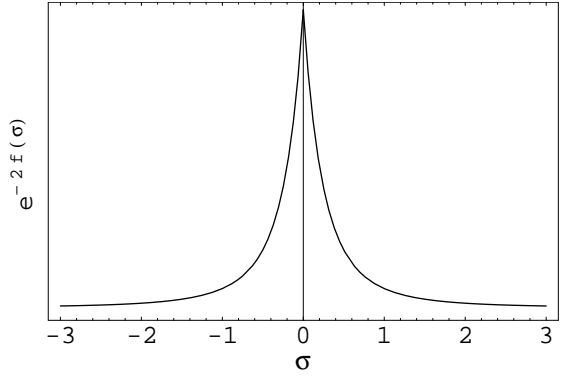}
\caption{The warp factor is a decaying function of the higher
  dimensional coordinate $\sigma$. We have used the same set of
  parameters as in the previous figure}
\end{figure}

The line element in this warped spacetime is given by 

\begin{equation}
ds^2 = e^{\frac{2 a_{1}}{k} e^{- k \vert \sigma \vert} }(\eta_{\mu\nu} 
dx^{\mu}dx^{\nu}) + d\sigma^2
\end{equation}

The Ricci scalar, R, for this background metric in the regions $\sigma \neq 
\sigma_j$ turns out to be : 

\begin{eqnarray}
R & = & - \left ( 8 a_{1} k  + 20 a_{1}^2 e^{- k
\vert \sigma \vert} \right ) e^{-k \vert \sigma \vert} 
\end{eqnarray}

Notice that R is always negative and approaches zero 
for $\sigma \rightarrow \pm \infty$. This 
implies that the spacetime is asymptotically flat.

It will be interesting to know the matter stress energy that gives
rise to such a bulk line element. The energy density and pressures 
(for $\sigma \neq \sigma_j$) are given as 

\begin{eqnarray}
\rho & = & - p_{i}  =  -(6 {a_{1}}^2 e^{- k \vert \sigma \vert} + 3
 a_{1} k )e^{-k \vert \sigma \vert}  \\
p_{\sigma} & = & 6 a_{1}^2 e^{-2 k \vert \sigma \vert} 
\end{eqnarray}

\begin{figure}[htb]
\includegraphics[width= 8cm,height=5cm]{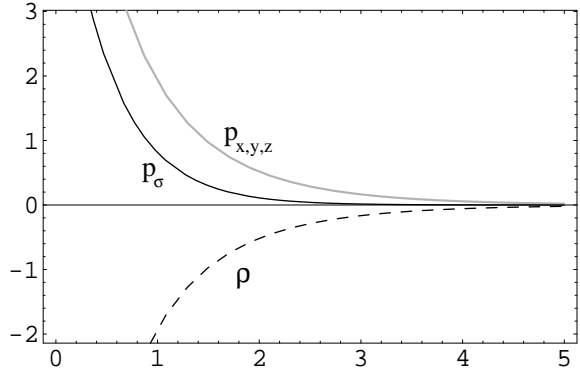}
\caption{The matter stress energy in the region $\sigma \neq \sigma_j$
and $\sigma >0$ is plotted with same set of parameters}
\end{figure}

for the range of extra dimensional coordinate excluding the location of the
brane(s). In Fig. (3) we have plotted the pressures and energy density as a
function of the transverse coordinate $\sigma$ (for $\sigma \neq \sigma_j$
and $\sigma >0$). For $\sigma <0$ the pressures and energy density are
the same because the functions are even in nature.  

It is worth mentioning that the matter stress-energy which acts as a
source for the warped geometry satisfies the SEC ($\rho + \sum p_{j} \ge 0$) 
though the WEC ($\rho \ge 0$) and NEC ($\rho + p_{j} \ge
0$) are violated  {\cite{ec}. 

\section{RS I two--brane model and the hierarchy problem}

%\subsection{Brane Set Up}

In RS I there are two branes located at $\sigma=0$ and $\sigma=\pi r_{c}$.
The extra dimension between the two 3--branes 
is defined on a $S_{1}/Z_{2}$ orbifold with branes located at the
orbifold fixed points $\sigma = 0, \pi r_{c}$. Let us now focus on the
equation (2.8) which reduces to the following form in the RSI set-up.  

\begin{equation}
3 f''  =  - \frac{1}{4 M^3} \frac{\alpha_{T} V(T) T'^2}{\sqrt{1 + T'^2}} +
\frac{\tau_{(1)}}{4 M^3} \delta(\sigma) +\frac{\tau_{(2)}}{4 M^3}
\delta(\sigma - \pi r_{c}) 
\end{equation}

where, M is the fundamental scale in five dimensions. The
second derivatives of the warp factor $ f(\sigma) = -\frac{a_1}{k}
e^{-k \vert \sigma \vert}$ (assuming periodicity of $f(\sigma)$) is given by :

\begin{eqnarray}
f''(\sigma) & = & -a_{1} k e^{- k \vert \sigma \vert} + 2 a_{1}
 e^{- k \vert \sigma \vert} \left[ \delta(\sigma) -
 \delta(\sigma - \pi r_{c})\right]
\end{eqnarray}

We assume that the bulk scalar field does not have any
interaction with on--brane matter. 
Comparing Eqn. (3.1) with Eqn. (3.2) we obtain the tension on the two branes as 

\begin{eqnarray} 
\tau_{(1)} & = & 24 M^3 a_{1} \\
\tau_{(2)} & = & - 24 M^3 a_{1} e^{- k \pi r_{c}}
\end{eqnarray}

It can be seen from the above relations that 
{\em unlike} the RS I case the brane tensions 
are {\em not equal} in magnitude. 
There is a scaling in the tension of the brane 
at $\sigma = \pi r_{c}$ which we consider as the visible
brane. The visible brane is a negative tension brane and its tension can be
chosen to be lower than that of the positive tension brane.
%\subsection{Solving the hierarchy problem} 
 
In order to address the hierarchy problem we 
integrate the bulk gravity action over the extra dimension and compare it
with the usual Einstein--Hilbert action for four dimensional gravity.
This results in the following relation between the four dimensional 
Planck scale ($M_{Pl}$)
and the fundamental five dimensional scale (M) : 

\begin{equation}
M_{Pl}^2 = \frac{2 M^3}{k} \left[ Ei \left (\frac{2 a_{1}}{k}
    \right) - Ei \left (\frac{2 a_{1}}{k} e^{ -k
    \pi r_{c}} \right)\right]
\end{equation}

where, ``Ei'' represents the exponential integral function {\cite{ei}}. 
For $\frac{a_1}{k}=39$ we obtain a solution of the hierarchy problem
for $k r_{c} = 5$. For such typical values one finds that a choice of fundamental 
(higher dimensional) scales of the TeV range for all parameters 
(such as five dimensional Planck scale, $m_0$ etc.) would reproduce the 
four dimensional Planck scale of $10^{19}$ GeV (using the relation in Eqn 3.5) 
leaving the masses on the visible brane in the TeV range. 
This shows that the heirarchy of scales
is a `derived' notion and there is no heirarchy of scales in a 
higher dimensional world.   
Note the reduction of the value of $kr_c$ as compared to the RS I
case. 
This is essentially due to
the modified nature of the warp factor in our solution here.
Following RS I, we can also show how the Higgs vacuum expectation
value and the masses of elementary particles will scale on the
negative tension brane ($m= m_0 e^{e^{-kr_c\pi}}$).

\section{RS II one--brane Model and localisation of fields}

In order to investigate whether the metric tensor
fluctuations give consistent four dimensional gravity we consider
the RS-II set up {\em{i.e.}} the transverse space is now infinitely
extended {\cite{rs}}. The regulator brane is now sent to infinity by making the
radius of the extra dimension to be infinite. The positive 
tension brane is now treated as our visible world. Let us consider a  
general linearized tensor fluctuation to the metric and check the
behavior of gravitational zero modes. 

\subsection{Gravity localisation}

We restrict ourselves for the metric fluctuation, 
$h_{\mu\nu} (x^{\mu}, \sigma)$ of the 4-dimensional world on the three
brane and consider the following metric :  

\begin{equation}
ds^2 = e^{- 2 f(\sigma)} (\eta_{\mu\nu} + h_{\mu\nu} (x^{\mu},\sigma)) dx^{\mu} dx^{\nu} 
+ d \sigma^{2}
\end{equation}

Keeping the background in the non-conformally flat form we proceed to 
analyse the gravitational fluctuation under RS gauge 
conditions - (i) $\partial_{\mu} h^{\mu}_{\nu} = 0$, (ii) $h^{\mu}_{\mu} = 0$.  
The linearized gravity fluctuation equation is then simply the covariant scalar
wave-equation. For a fluctuation of the form 
$h_{\mu\nu}(x^{\mu}, \sigma) = \phi(\sigma) \bar h_{\mu\nu}(x^{\mu})$ 
with $ \nabla^{\rho} \nabla_{\rho}{\bar h_{\mu\nu}} = m^2 \bar
h_{\mu\nu}$ , m being the four dimensional mass of Kaluza-Klein mode we 
perform the mode analysis {\cite{thick,gravloc}}. The equation
satisfied by $\phi(\sigma)$ always admits a zero mode ({\em {i.e.}} m
= 0) solution, $\phi_{0}(\sigma) =
constant$. Now, inserting this solution to the action {\cite{gravloc}}
from which the gravity fluctuation equation is derived one obtains the
following normalisability condition for the zero-mode wave function  

\begin{equation}
\int{e^{- 2 f(\sigma)} d\sigma} < \infty
\end{equation}

Using the expression of $f(\sigma)$ obtained in equation (2.9) we
find that the integrand is finite over the entire domain of the extra
dimension, from which we conclude that the above condition is
satisfied in our case. Therefore it can be said that gravity is 
localised around the brane with the zero mode being the massless
graviton.

\subsection{Fermion Localisation} 

We now turn our attention to the localization of spinor fields on the brane. 
The bulk fermion coupled to the scalar field in 5D gives rise to two
chiral fermionic zero modes in four dimensions. Depending on the 
nature of the coupling one of these two modes 
is found to be localized on the brane while the other is not
{\cite{loc,fermiloc}}. In the background of the scalar field 
coupled to gravity we find that the left chiral modes only can be
confined to the brane for a positive Yukawa coupling. 

The Lagrangian for a Dirac fermion propagating in the five dimensional warped 
space with the metric (2.5) is :

\begin{equation}
\sqrt{-g} {\cal{L}}_{Dirac} = \sqrt{-g}\hspace{.03in}(i \bar{ \Psi} \Gamma^{M} {\cal{D}}_{M} \Psi
- \eta_{F} \bar{ \Psi} \mbox{F}(T) \Psi )
\end{equation}
 
The matrices $\Gamma^{a} = (e^{f(\sigma)} \gamma^{\mu}, -i \gamma^{5})$ provide a 
four dimensional
representation of the Dirac matrices in five dimensional curved space. Where $\gamma^{\mu}$
and $\gamma^{5}$ are the usual four dimensional Dirac matrices in chiral representation.

The Dirac Lagrangian in 5D curved spacetime then reduces to the following form
 
\begin{equation}
\sqrt{-g} {\cal{L}}_{Dirac} = e^{- 4 f (\sigma)}  \bar{\Psi}\left[i  e^{f(\sigma)}\gamma^{\mu}
\partial_{\mu} + \gamma^{5} (\partial_{4} -2 f'(\sigma)) - \eta_{F} \mbox{F}(T) \right ] \Psi  
\end{equation}

The dimensional reduction from 5D to 4D is performed in such a way that the
standard four dimensional chiral particle theory is recovered. 
Separating the variables the five dimensional spinor can be written as: $\Psi(x^{\mu},\sigma) =
\Psi(x^{\mu}) \xi(\sigma)$. Since the four dimensional massive
fermions require both the left and right chiralities
it is convenient to organise the spinors with respect to
$\Psi_{L}$ and $\Psi_{R}$ which represent four component spinors living in five
dimensions given by $\Psi_{L,R} = \frac{1}{2} (1 \mp \gamma_{5})\Psi$.
Hence the full 5D spinor can be split up in the following way

\begin{equation}
\Psi(x^{\mu},\sigma) = \left( \Psi_{L}(x^{\mu})\xi_{L}(\sigma) +
\Psi_{R}(x^{\mu})\xi_{R}(\sigma) \right)
\end{equation}

where $\xi_{L.R}(\sigma)$
satisfy the following eigenvalue equations

\begin{eqnarray}
 e^{-f(\sigma)}\left [\partial_{\sigma}-2 f'(\sigma) - \eta_{F}   \mbox{F}(T) \right ] 
\xi_{R}(\sigma)
& = & -m \xi_{L}(\sigma) \\
 e^{-f(\sigma)} \left [\partial_{\sigma}-2 f'(\sigma) + \eta_{F}   \mbox{F}(T) \right ] 
\xi_{L}(\sigma)
& = & m \xi_{R}(\sigma)
\end{eqnarray}

Here $m$ denotes the mass of the four dimensional fermions.
The full 5D action reduces to the standard four dimensional action 
for the massive chiral fermions, 
when integrated over the extra dimension \cite{fermiloc}, if
(a) the above equations are satisfied by the bulk 
fermions and (b) the following orthonormality conditions are obeyed.

\begin{eqnarray}
\int_{-\infty}^{\infty} e^{-3 f(\sigma)} \xi_{L_{m}} \xi_{L_{n}} d\sigma =
\int_{-\infty}^{\infty} e^{-3 f(\sigma)} \xi_{R_{m}} \xi_{R_{n}} d\sigma = \delta_{m n} \\
\int_{-\infty}^{\infty} e^{-3 f(\sigma)} \xi_{L_{m}} \xi_{R_{n}} d\sigma = 0
\end{eqnarray}

The massless (i.e. $m = 0$) fermions for a Yukawa coupling of the form
$F(T) = T$ are represented by the following wave functions

\begin{eqnarray}
\xi_{L} = exp \left[ \frac{2}{k} \left(- a_{1} e^{- k \vert \sigma \vert} - \eta_{F}
  \sqrt{\frac{2}{a_{1} k}} e^{\frac{k}{2} \vert \sigma \vert} \right) \right]
  \\
\xi_{R} = exp \left[ \frac{2}{k} \left(- a_{1} e^{- k \vert \sigma \vert} + \eta_{F}
  \sqrt{\frac{2}{a_{1} k}} e^{\frac{k}{2} \vert \sigma \vert} \right) \right] 
\end{eqnarray}

If the coupling constant, $\eta_{F} = 0$, it is clear from the above expressions that both
the left and right chiral modes decay away from the brane. If
$\eta_{F} > 0$ then $\xi_{L}$ decays but $\xi_{R}$ grows and the
reverse phenomena takes place for $\eta_{F} < 0$.  

\begin{figure}[htb]
\includegraphics[width= 8cm,height=5cm]{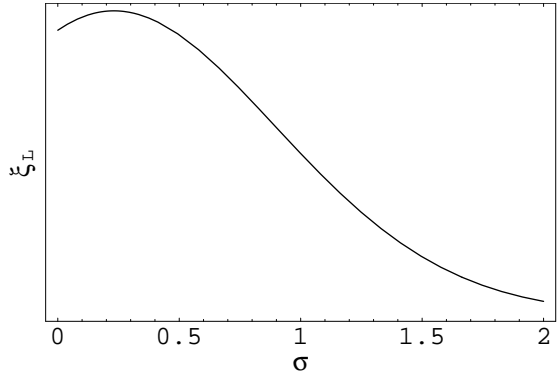}
\caption{The left chiral zero modes are localised around the brane. We
have used unit coupling constant and kept the set of parameters unchanged}
\end{figure}

In Fig.(4) we have shown the nature of variation of the left chiral
mode w.r.t. the extra dimension. The left chiral mode
is localised around the brane for positive Yukawa coupling 
whereas for an opposite Yukawa coupling
we get the right chiral modes to be confined around the brane. 

\section{Conclusion}

We now summarize below the results obtained.
 
(i) We have considered the tachyon condensate as a bulk scalar 
coupled to gravity. 
Exact analytical expressions for the warp factor
and scalar field have been obtained from the solution of the full
Einstein-scalar equations. The nature of the scalar field
potential depends on the sign of the coupling constant
$\alpha_{T}$. The background geometry obtained in this case is however
described by a negative, non-constant Ricci scalar with delta function
singularity at the location of the brane; it goes to zero
asymptotically. The stress-energy tensor,  giving rise to this geometry,
violates the null and weak energy conditions whereas it satisfies the
strong energy condition.    

(ii) The resolution of the gauge hierarchy problem is attempted in
this background geometry for a RS I type model consisting of two branes 
situated 
at the two fixed points in an $S^{1}/Z_{2}$ orbifold. We find that the brane
tensions are of unequal and opposite magnitude. 
The relation between the five dimensional fundamental mass scale 
and the four dimensional Planck scale can be brought to same order for a 
particular choice of the model parameters. We also point out how
the Higgs expectation value and the masses scale on the negative
tension brane.  

(iii) 
The graviton zero mode in this background geometry is found to
localised and normalisable. Additionally,
fermionic zero modes with left chirality are found to be localised
around the brane in the presence of a Yukawa coupling between the scalar  
and spinor field. The right chiral modes may be found on the brane for
an opposite value of the Yukawa coupling parameter.   

The fate of the massive graviton and fermion modes in this
background geometry is an obvious next question we need to answer. 
Furthermore, we also would like to know about cosmological solutions
on the brane by assuming a flat FRW on--brane line element. We hope 
to come back with these issues and other related ones in future.

\section*{Acknowledgments}
RK thanks IIT Kharagpur for support. 
 
\maketitle
\vspace{.2in}

%%%%%%%%%%%%%%%%%%%%%%%%%%%%%%%%%%%%%%%%%%%%%%%%%%%%%%%%%%%%%%%%%%%%%%%%%%%%%%%%%%%%%%%%%%%%%%%%%%%%%%%%%%%


\begin{references}


\bibitem{kk} T. Kaluza, Sitzungsber. Preuss. Akad. Wiss., Phys-Math.Kl.,
Berlin Math. Phys., Bd. K1 966 (1921) ; O. Klein, Z. Phys. {\bf 37}, 895
(1926); A. Salam and J. Strathdee, Ann. Phys. {\bf 141}, 316 (1982);
M. J. Duff, B.E.W. Nilsson and C. N. Pope, Phys. Rep. {\bf 130 C}, 1
(1986); T. Appelquist, A. Chodos and P. G. O. Freund, 
{\em Modern Kaluza--Klein Theories}, Reading, MA, Addison--Wesley (1987).

\bibitem{string} M. S. Green, J. H. Schwarz and E. Witten, {\em
Superstring theory}, Cambridge University Press (1987); J. Polchinsky,
{\em String theory}, Cambridge University Press (1997);

\bibitem{rs} L. Randall and R. Sundrum, Phys. Rev. Lett. {\bf 83}, 3370
(1999); {\em ibid} Phys. Rev. Lett. {\bf 83}, 4690 (1999)

\bibitem{modrs} W. D. Goldberger and M. B. Wise, Phys. Rev. Lett. {\bf
  {83}}, 4922 (1999); R. Gregory, V. A. Rubakov and S. M. Sibiryakov,
  Phys. Rev. Lett. {\bf 84}, 5928 (2000); G. Dvali, G. Gabadadze and
  M. Porrati, Phys. Lett. {\bf B484}, 112, (2000) 

\bibitem{radion}
P. J. Steinhardt, Phys. Lett. {\bf{B 462}} 41, (1999); C. Csaki,
M. Graessser, L. Randall nad J. Terning, Phys. Rev. {\bf D 62} 045015
(2000); C. Charmousis, R. Gregory and V. A. Rubakov,
Phys. Rev. {\bf{D 62}} 067505 (2000); P. Brax, C. Bruck and A. C. Davis and
C. S. Rhodes, Phys. lett. {\bf{B 531 }}, 135, (2002)  

\bibitem{thick} 
C. Csaki, J. Erlich, C. Grojean and T. Hollowood,  Nucl. Phys. {\bf B 584}, 359 (2000) ;  
O. DeWolfe, D. Z. Friedman. S. S. Gubser and A. Karch, Phys. Rev. {\bf
  D 62} 046008 (2000); 
C. Csaki, J. Erlich, T. Hollowood and Y. Shirman, Nucl. Phys. {\bf
  B 581}, 309 (2000) 
H. Davoudiasl, J.L. Hewett and T.G. Rizzo, Phys.Lett. {\bf B 473} 43 (2000);
S. C. Davis, JHEP {\bf 0203}, 054 (2002);
E. E. Flanagan, S.H. Henry Tye and I. Wasserman, Phys.Lett. {\bf B 522} 155 (2001);



\bibitem{loc} B. Bajc and G. Gabadadze, Phys. Lett. {\bf B 474}, 282
 (2000); 

\bibitem{gravloc} J. Garriga and T. Tanaka,  Phys. Rev. Lett. {\bf 84} 2778 (2000); 
J. Lykken and L. Randall, {JHEP} {\bf 0006}, 014, (2000); 
A. Karch and L.Randall, {JHEP} {\bf 0105}, 017, (2001); 
S. B. Giddings, E. Katz and L. Randall,  {JHEP} 0003 (2000); 
I. Brevik, K. Ghoroku, S. D. Odintsov and M. Yahiro,  Phys.Rev. {\bf D66}, 064016 (2002)

\bibitem{fermiloc} S. L. Dubovsky, V. A. Rubakov and P. G. Tinyakov,
 Phys. Rev. {\bf D 62}, 105011 (2000) 
 R. Jackiw and C. Rebbi, Phys. Rev. {\bf D 13}, 3398 (1976); Y. Grossman and 
N. Neubert, Phys. Lett. {\bf B 474} 361 (2000)
R. Jackiw and C. Rebbi, Phys. Rev. {\bf D 13}, 3398 (1976); Y. Grossman and 
N. Neubert, Phys. Lett. {\bf B 474} 361 (2000);
C. Ringeval, P. Peter, J. P. Uzan, Phys. Rev. {\bf D 65}, 044416 (2002);
S. Ichinose, Phys.Rev. {\bf D 66}, 104015 (2002)

\bibitem{cosmo} D. Langlois, gr-qc/0102007 (Proceedings of the
9th Marcel Grossmann meeting, July 2000 (Rome); gr-qc/0205004 (Proceedings
of Journees Relativistes, Dublin 2001 and references therein
; E. Flanagan, S.H. Henry Tye and I. Wasserman,
Phys. Rev. {\bf D 62}, 024011 (2000); H. Stoica, S. H. Henry Tye
and I. Wasserman, Phys. Letts. {\bf B 482}, 205 (2000); J. S. Alcaniz,  Phys. Rev. {\bf D65}, 123514, (2002)

\bibitem{particle} M. Besancon, hep-ph/0106165 ; Y. A. Kubyshin, hep-ph/0111027
and references therein.

\bibitem{sen} A. Sen, { JHEP} {\bf 0204}, 048, (2002); {\em{ibid}} { JHEP}
  {\bf 0207}, 065 (2002); {\em{ibid}} 
  Mod. Phys. Letts. {\bf{A}} {\bf{ 17}}, 1797 (2002); 

\bibitem{bi} G. W. Gibbons, K. Hori and P. Yi, Nucl. Phys. {\bf{B 596}} 
(2001) 136; A. Sen, J. Math. Phys {\bf{42}} (2001) 2844.

\bibitem{tachyon} M. R. Garousi, Nucl. Phys. {\bf{B 584}}, 284 (2000);
  J. Kluson, Phys Rev. {\bf{D 62}}; E. A. Bergshoeff, M. de Roo,
  T. C. de Witt, E. Eyreas and S. Panda, JHEP {\bf{0005}}, 009 (2000);

\bibitem{tachycosmo} G.W. Gibbons, Phys Letts. {\bf {B 537}}, 1
  (2002); M. Fairbairn and M. H. G. Tytgat, Phys. Lett. {\bf{B 546}},
  1, (2002); G. Shiu and I. Wasserman, Phy. Lett. {\bf{B 541}}, 6, (2002)  
 

\bibitem{newbulk} B. Mukhopadhyaya, S. Sen, S. SenGupta,  Phys.Rev. {\bf D 65} 124021 (2002);  
M. Gogberashvili, Phys.Lett. {\bf B 553}, 284 (2003); 
R. Koley and S. Kar,  Mod. Phys. Letts. {\bf{A}} {\bf{20}}, 363
(2005) ;     
  K. Ghoroku, hep-th/0402102; 
         
\bibitem{ec} For a useful summary on the details of the energy conditions 
see {\em Lorentzian Wormholes : from Einstein to Hawking} by Matt Visser (AIP, 1995)

\bibitem{ei} M. Abrmowitz and C. A. Stegan, {\em{Handbook of
    Mathematical Functions with Formulas, Graph, and Mathematical
    Tables}} (Dover, New York, 1972)




\end{references}
\end{document}